\documentclass{emulateapj}
\usepackage{graphicx}
\begin{document}

\title{Collapse of Neutron Stars to Black Holes in Binary Systems: 
A Model for Short Gamma Ray Bursts}

\author{Charles D. Dermer\altaffilmark{1}
\&
Armen Atoyan\altaffilmark{2}}

\altaffiltext{1}{E. O. Hulburt Center for Space Research, Code 7653
Naval Research Laboratory, Washington, D.C. 20375-5352; 
\small dermer@gamma.nrl.navy.mil}
\altaffiltext{2}{CRM, Universit\'e de Montr\'eal, Montr\'eal, Canada H3C 3J7;
\small atoyan@crm.umontreal.ca}

\begin{abstract}
The accretion of $\approx 0.1$ -- 1 $M_\odot$ of material by a neutron
star through Roche lobe overflow of its companion or through
white-dwarf/neutron-star coalescence in a low mass binary system could
be enough to exceed the critical mass of a neutron star and trigger
its collapse to a black hole, leading to the production of a short
gamma-ray burst (SGRB). In this model, SGRBs would often be found in
early-type galaxies or in globular cluster environments, though they
could also be formed through stellar-wind accretion in high-mass
binary systems of star-forming galaxies.  Although the collapse event
is likely to be of short ($\ll 1$ s) duration, afterglow complexity
could be produced by neutron decoupling in the fireball and subsequent
accretion of the companion remnant, for example, the disrupted
white-dwarf core.
\end{abstract}


\section{\label{intro}Introduction}

Recent discoveries from Swift and HETE-II have led to the
identification of host galaxies for short GRBs (SGRBs), which is the
class of GRBs with durations $\lesssim 2$ s that tend to be spectrally
harder than the classical long-duration GRBs
\citep[LGRB;][]{kou93}. GRB 050509B, the first rapidly localized SGRB
\citep{geh05}, displayed a single short spike with duration of 40$\pm
4$ ms. No radio or optical counterpart emission was detected. A large
elliptical cD galaxy at redshift $z = 0.225$ overlapped the
9.3$^{\prime\prime}$ error radius of the XRT on Swift, and is the
likely host galaxy \citep{blo06}. No supernova emissions were detected
to faint limits, ruling out an origin associated with supernovae,
including Type 1a SNe \citep{hjo05,kul05}. Its apparent isotropic
energy was $\approx 4.5\times 10^{48}$ ergs, with a smaller absolute
energy release if the emission was collimated.

In contrast, GRB 050709, a SGRB discovered with HETE-II, is associated
with an irregular star-forming galaxy at $z = 0.160$
\citep{fox05}. This burst displayed a 70 ms single-spiked 30 -- 400
keV hard pulse followed by a 130 s long smooth bump \citep{vil05}. The
energy released in the bump was about the same as the energy released
in the spike.  X-ray flares occurred $\approx 100$ s and $\approx 16$
days after the GRB. The afterglow was also detected at
optical frequencies, with a jet break apparent at 10 days after the
burst. A beaming factor $f_b \approx 0.03$ implies an absolute energy
release of $\approx 2\times 10^{48}$ ergs \citep{fox05}.

GRB 050724 is a Swift-detected SGRB that exhibited a 250 ms spike
followed by an $\approx 3$ s 15 -- 150 keV hard bump and an $\approx
100\,$s 15 -- 25 keV soft bump starting $\approx 1\,$s and $\approx
40\,$s after the trigger, respectively \citep{bar05}. This burst is
clearly associated with an elliptical galaxy at $z = 0.258$. The X-ray
afterglow of GRB 050724 is remarkable in that it exhibits a rapid
decline in the X-ray flux at $\approx 200$ -- 400 s after the
trigger---similar to the behavior seen in LGRBs
\citep{tag05}---and an X-ray flare peaking $\approx 1/2$ day later. A
break in the optical light curve $\approx 1$ day after the GRB
suggests $f_b\approx 0.01$ \citep{ber05}, implying an absolute total
energy of $\approx 4\times 10^{48}$ ergs. The energy released in the
soft bump and the later X-ray flare is comparable to the energy
measured in the prompt phase.

Besides these bursts, two other SGRBs have also been rapidly
localized.  GRB 050813 is associated with a cluster elliptical galaxy
at either $z = 0.722$ or $z \gtrsim 1$ \citep{berger05,gla05}. GRB
050906 is possibly associated with a nearby galaxy at $z \cong 0.03$
\citep{lt05}. The association of 3 out of $5$ SGRBs with elliptical
galaxies excludes models associated with massive stars as a common
origin for these sources \citep{pro06}. \citet{gal05} perform an
archival analysis of SGRBS localized through the Interplanetary
Network to enlarge the sample to include two sources at $z < 0.15$ and
two others at $z
\gtrsim 0.3$.

Here we consider whether the collapse of a neutron star (NS) to a
black hole (BH) in a binary system, as considered earlier for LGRBs by
\citet{vs99} and for SGRBs by \citet{mrz05}, either through
accretion-induced collapse (AIC) or through the coalescence of the
low-mass companion with the NS, can produce SGRBs.  Statistical
arguments for the model are presented in Section 2, explanations for
X-ray flares are considered in Section 3, and a discussion and summary
are given in Section 4.

\section{Statistics of Neutron Stars in Binaries}

Observations of low mass X-ray binaries (LMXBs) and high-mass XBs
(HMXBs) in the Galaxy with both NS and BH companions provide support
for the idea that the collapse of NSs to BHs could occur in binary
systems. The most widely considered evolutionary pathway to form a BH
binary (BHB) involves instead the collapse of the evolved core of a
massive ($\approx 25$ -- 100 $M_\odot$) star in a primordial binary
directly to a BH with mass $\approx 10 ~M_\odot$ \citep[for review,
see][]{lew95}.  Depending on the range of initial stellar masses that
collapse to NSs and BHs, the rate of production of LMNSBs in the
Galaxy is $ 10^{-5}\lambda_{-5}$ yr$^{-1}$, with $\lambda_{-5} \sim 1$
\citep{pz97}, at least by a factor $\approx 30$ greater than the
formation rate of LMBHBs.  This should lead to a preponderance of
NSXBs over BHXBs in our Galaxy, contrary to observation. If the mean
lifetime of NSXBs is smaller than BHXBs, then this discrepancy could
be resolved. One possibility is that a large fraction of NSXBs in
low-mass systems form BHXBs through AIC. In this case $\approx 10^4$
-- $10^5$ such events should occur over the lifetime of the Galaxy.
Another possibility is that the number of NSXBs could be reduced
relative to BHXBs if the rate of coalescence events, limited by the
measured rate of SGRBs, is comparable to the LMB formation rate.

Approximately $ 25$\% of BATSE GRBs are SGRBs \citep{kou93}, implying
that SGRBs occur at the rate
\begin{equation}
\Lambda_{SGRB}\gtrsim {5000 \over (f_b/0.03)}\;{\rm~yr}^{-1}
\label{LSGRB}
\end{equation}
throughout the universe for a full-sky BATSE GRB rate of 550 yr$^{-1}$
\citep{ban02}, where $f_b$ is the mean beaming factor of SGRBs. Swift
is less sensitive to SGRBs than BATSE, and detects only 1 SGRB every 2
-- 3 months \citep{bar05}, or $\lesssim 10$\% of the Swift GRBs.  The
present data are inadequate to determine the mean redshift $\bar
z_{SGRB}$ of Swift SGRBs, which may, moreover, be different from the
mean redshift of BATSE SGRBs.  The value $\langle
V/V_{max}\rangle_{SGRB} = 0.39 \pm 0.02$ measuring the deviation from
a Euclidean distribution for BATSE SGRBs is significantly larger than
for BATSE LGRBs \citep[$\langle V/V_{max}\rangle_{LGRB} = 0.29 \pm
0.01$;][and references therein]{gp05}, but is also sufficiently
different from 0.5 to suggest that cosmological effects are starting
to play a role.

The total number of LMXBs in galaxies of different types is poorly
known, because LMXBs may be quiescent for most of their lifetime.
Measurements \citep{gil04} show that, within a factor of $\approx 2$,
the number of luminous ($\gtrsim 3\times 10^{35}$ ergs s$^{-1}$) LMXBs
is proportional to the stellar mass of the host galaxy for spirals and
ellipticals. If the total number of LMXBs is proportional to the
number of luminous LMXBs, then we can estimate the production rate
$\Lambda_{LMXB}$ of LMXBs in the local universe using the LMNS binary
production rate in our Galaxy per unit stellar mass implied by the
calculation of \citet{pz97}.  The result for $z \lesssim 1$, using a
stellar mass in the Milky Way of $3\times 10^{11} M_\odot$, is
\begin{equation}
\Lambda_{LMXB}(\leq z)  \; \simeq \; 8\times 10^3 
\;z^3\;({\rho_*\over 0.1}) \lambda_{-5}  ~{\rm yr}^{-1}\;,
\label{Lambda}
\end{equation}
where $\rho_*$ \citep[$\approx 0.1$; see][]{rud03} is the fraction of
baryon mass in the form of stars in the local universe, the ratio of
baryon density compared to the critical density is 5\%, and we use a
Hubble constant of $ 70$ km s$^{-1}$ Mpc$^{-1}$.  This rate exceeds
the SGRB rate, eq.\ (\ref{LSGRB}), if
\begin{equation}
z \gtrsim 0.85 (\;{f_b \over 0.03}\;{\rho_*\over 0.1}
\;\lambda_{-5}\; \epsilon \;)^{-1/3}\;,
\label{z}
\end{equation}
where $\epsilon~ (\leq 1)$ is the fraction of NSs in LMBs that
collapse to a BHs over the $\approx 10^{10}$ yr lifetime of
galaxies. Unless $\lambda_{-5} \gg 1$, a significant fraction of NSs
in binaries must therefore collapse to BHs to provide the observed
rate of SGRBs.

A contact or short-period LMXB forms through spiral-in of the lower
initial mass secondary during a common envelope phase while the higher
initial mass primary is evolving through a giant phase.  If the
primary evolves to form a NS before the secondary coalesces with the
helium core of the primary (which depends on the initial separation
and initial masses of the stars), then a close LMNS binary will
form. When the secondary evolves through its hydrogen-burning phase
and expands to overfill its Roche lobe \citep{pod02}, additional
accretion onto the NS can trigger collapse to a BH and the formation
of a SGRB. Alternately, the secondary can lose angular momentum
through gravitational radiation and magnetic braking to spiral in and
coalesce with the NS.  If the pulsar had evaporated the companion, all
that may remain of the secondary is a helium core.

Formation of a BH through AIC or white-dwarf/NS coalescence requires
that the NS accretes enough mass to exceed the maximum stable mass of
a NS.  Depending on the equation of state of nuclear matter, this mass
is $\approx 2.2$ -- $2.9 M_\odot$ \citep{kb96}, though rotation could
increase these values by $\lesssim 20$\% \citep{st83}. Sufficient mass
could also be accreted from stellar winds to trigger AIC in
HMBs. These systems would be found only in star-forming galaxies, and
may display distinctive signatures due to the different
environment.

Recent observations show that NSs with masses significantly exceeding
the Chandrasekhar mass are found in LMBs.  PSR J0751+1807
is a 2.1 $M_\odot$ millisecond pulsar in a 0.26 d orbit around an
$\approx 0.2 M_\odot$ helium white dwarf companion \citep{nic05}.  As
its orbit decays, the white dwarf will overflow its Roche lobe.  A
number of evolutionary outcomes are possible, including an
ultracompact X-ray binary phase and tidal disruption of the white
dwarf \citep{elc97}. Given the large mass of the NS, the NS may also
collapse to a BH when it exceeds its maximum mass through
accretion. NS masses $\gtrsim 1.7 M_\odot$ are found for several
millisecond pulsars in the globular cluster Terzan 2 \citep{ran05}.
The masses of most millisecond pulsars are, however, near the
Chandresakhar limit, even allowing for the mass needed to spin up the
pulsar \citep{jac05}, but selection biases may exist against the
detection of high mass NSs in millisecond pulsar systems if magnetic
field decay accompanies accretion.

The total mass accreted by the compact companion in LMXBs over the
lifetime of the Galaxy is $\approx 10^5 M_\odot$, using the LMXB
energy output computed by \citet[][eq.\ (19)]{gil04}, for a 10\%
accretion radiation efficiency and a bolometric-correction factor of
$\approx 2$. Most of this energy is radiated by sources with
luminosity $\lesssim 2\times 10^{38}$ ergs s$^{-1}$, the Eddington
limit of a 1.4 $M_\odot$ object.  If NSs in $\sim 10^5$ LMNSBs each
accrete about the same fraction of this mass during that time, then a
large fraction will collapse to BHs in the Galaxy. The proportionality
of the number and luminosities of LMXBs in systems with old stellar
populations compared to the Milky Way indicates that this mechanism
will also operate effectively in early-type galaxies.  The larger
fraction of LMXBs in globular clusters, which are more prevalent in
elliptical galaxies \citep{sar03}, and the long time required for AIC,
means that SGRBs would often be found at large distances from the
center of E/S0 galaxies.

\section{X-ray Flares from SGRBs}

One of the surprises of the Swift and HETE-II results is the complex
X-ray afterglow behavior of SGRBs. It is hard to invoke density
inhomogeneities for bright flares observed hundreds of seconds to days
or more after the trigger, especially in those cases where SGRBs are
associated with old stellar populations that should inhabit a rarefied
quasi-uniform ISM.

There are, however, at least four other processes that can produce
episodes of emission from SGRBs on different timescales. First is the
duration of engine activity. Before reaching the moment of collapse,
the NS in a LMXB could have accreted enough mass and angular momentum
to have been spun up to millisecond periods and would, in effect, be a
recycled pulsar.  As in the case of a hypermassive NS formed in binary
neutron-star coalescence \citep{shi06}, the newly formed black hole
would be surrounded by a dense torus that accretes on subsecond
timescales, accounting for the first burst of emission of the SGRB.
Millisecond structure in the lightcurve could be explained by
angular-momentum transport through the magneto-rotational instability
in the material accreting on the newly formed BH.

A second timescale that could appear in the X-ray lightcurve of SGRBs
relates to the deceleration of the blast wave as it sweeps up material
from the ISM. For an apparent isotropic energy release of $E_0 =
10^{50}E_{50}$ ergs from a blast wave with mean entropy per baryon
$\Gamma_0 = 300 \Gamma_{300}$ that decelerates in a uniform
surrounding medium with density $ n = 10^{-2}n_{-2}$ cm$^{-3}$, the
deceleration timescale
\begin{equation}
t_d = (1+z) \big({3E_0\over 4\pi n m_pc^5 \Gamma_0^8})^{1/3}
\cong 10 \big({E_{50}\over n_{-2} \Gamma_{300}^8}\big)^{1/3}\;{\rm s} \;.
\label{td}
\end{equation}
The value of $t_d$ represents the time when unabsorbed synchrotron
emission from a decelerating blast wave reaches its maximum value. The
value of $\Gamma_0$ is unknown for SGRBs, so the afterglow synchrotron
emission could explain emission peaks on various
timescales,\footnote{If SGRBs are sources of 100 MeV -- GeV radiation,
then GLAST will be able to constrain lower limits of $\Gamma_0$ from
$\gamma\gamma$ transparency arguments.} for example, emission bumps
observed a few seconds to minutes after the start of GRBs 050709 and
050724.  The peaking of this emission could however be concealed by
the decay phase of the prompt emission if $t_d$ is very short, which
could happen if $\Gamma_0 \gg 300$.

Another emission episode can occur because of neutron decoupling from
the fireball during its expansion phase \citep{dkk99,bm00}. The baryon load
would be rich in neutrons from the accreting neutron-star material.  If the
elastic nuclear interaction timescale is longer than the comoving
time, neutron decoupling occurs when $\Gamma_0 /\Gamma_{dcp}\equiv u~
(> 1)$, where the decoupling Lorentz factor $$\Gamma_{dcp} \cong (\ln
2 \Gamma_{dcp} )^{1/4}\;\big[\; {E_0 \sigma_{pn} \over (1+y) 4\pi
m_pc^2 \Delta^2}\;]^{1/4}\;$$
\begin{equation}
  \cong 150 \;[1 + 0.17\ln ({\Gamma_{dcp}\over 150} )]^{1/4}
\;\big[\; {E_{50} \over {(1+y)\over 2 }\Delta_9^2}\;]^{1/4}\;.
\label{Gdcp}
\end{equation}
Here $\sigma_{pn} \cong 30$ mb, $y$ is the ratio of neutrons to
protons in the expanding fireball, and $\Delta/c $ is the
characteristic timescale during which energy is injected, so that
$\Delta/c = 10^9\Delta_9/c = 33 \Delta_9$ ms.

The Lorentz factor of neutrons when they decouple is $\Gamma_n =
\Gamma_0 u^{-4/3}$, and the fraction of total energy in the decoupled
neutrons is $(1+ u^{4/3}/y)^{-1}$.  The neutrons will decay before
colliding with the forward proton shell when the comoving time
$t^\prime_n \cong x_{col}/\Gamma_n c \gg 900$ s, the mean neutron
lifetime. The collision radius where the shell of neutron-decay
protons impact the decelerating forward proton shell is $x_{col} =
2^{1/3} u^{8/9}x_d$, where $x_d = \Gamma_0^2 c t_d/(1+z) \cong
2.6\times 10^{16} (E_{50}/n_{-2}\Gamma_{300}^2)^{1/3}$ cm is the
deceleration radius.  If $x_d \gg 6\times 10^{15}\Gamma_{300}
u^{-20/9}$ cm, then most neutrons will decay before the impact.

From this, we derive the observer time of the collision, given by
$t_{col}\cong 2^{1/3} u^{32/9} t_d$.  When $u \cong 2$, a
neutron-decay proton shell with significant amount of energy will
collide with the forward shell, resulting in a strong shock that could
be observed as the second peak near 100 s in GRB 050724.  After the
shock crosses the shell and injection ceases, the flux rapidly
declines. This results from fast synchrotron cooling and expansion of
the shell. Note that only one flare can be produced from this process,
though other neutron decoupling scenarios can be considered
\citep{bel03}.

The preceding three mechanisms may also equally operate for NS/NS and
NS/BH mergers.  In the case of AIC of a NS to a BH, however,
the impact of gravitational and neutrino radiation produced 
at the time of collapse of the neutron star to a BH would heat 
and disrupt the companion star, causing the ejection of material 
outside the Roche surface.  Fallback of this
material onto newly formed black hole can trigger flares hours or days
later, as suggested by simulations of white dwarf mergers with BHs
\citep{fwh99}.  The stellar companion can also be disrupted through
tidal forces if its orbit is perturbed.  Only $\approx 10^{-5}
M_\odot$ of matter is required for X-ray flares from SGRBs, and
fallback of the disrupted stellar material could make flares weeks to
months after the event. We expect that a large fraction of BHs in
LMBHBs are just above the maximum stable mass of a neutron star if
this model is correct.

\citet{mrz05} also argue that shock interaction of the ejecta
with a red giant companion could make flares seen minutes after
the prompt emission. Such a process is unlikely to explain the 
rapid decline in GRB 050724 due to light travel-time delays
from different parts of the red-giant surface. 
Moreover, because comparable X-ray energy is detected
in the prompt emission and in the delayed X-ray flares, considerably more
energy must be directed toward the red giant than is emitted in the
beamed prompt emission. X-ray flares unassociated 
with SGRBs would be therefore be predicted from the unbeamed
shock-heated red giant surface.


\section{Discussion and Summary}

Although we have considered a model of SGRBs from the collapse of NSs
to BHs in binaries, as first proposed for SGRBs by \citet{mrz05}, 
SGRBs may represent a heterogeneous class.  The
currently favored scenario for SGRBs is
compact-object coalescence involving NS/NS and NS/BH binaries
\citep[brief reviews are given by][see \citet{rj01}
for Newtonian calculations and \citet{shi05} for fully general
relativistic calculations]{zm04,pir05}.  Such systems that will
coalesce within a Hubble time are known to exist, and can release
sufficient energy through $\nu\bar\nu \rightarrow e^+ e^-$ processes
\citep{jan99,lee05} to power a SGRB. Compact object merger scenarios
with short delay times may, however, conflict with the statistics of
SGRB host galaxy types and the properties of observed NS binaries
\citep{gal05,ngf06}.  Evolutionary calculations of the timescale for AIC of
NSs in LMXBs are required to determine if the model proposed here is
in accord with the preference of SGRBs for early-type galaxies.

The features that make compact-object coalescence an attractive model
for SGRBs also apply here, because both involve bursting events
derived from the gravitational energy released by the collapse of a
supramassive NS to a BH.  Initial loss of angular momentum support
along the rotation axis of the NS would form evacuated axial regions
and a massive, dense torus of neutron-star material that would accrete
on ms timescales \citep{vs98,shi06}.  Along the poles, a baryon-dilute
jet would be created through $\nu\bar\nu$ annihilation \citep{ajm05}
or MHD \citep{ds02} processes.  Insofar as the progenitor systems
would not involve stellar winds or supernovae, the collimated outflows
would be launched into a low-density external medium.


The occurrence of flaring emissions long after the SGRB takes place
was not, however, predicted in NS/NS or NS/BH coalescence scenarios.
For NS/NS mergers, the collapse is over after only a few tens of ms
\citep{jan99}.  The torus formed by a transient highly magnetized
hypermassive NS collapses to a BH formed in a NS/NS merger lasts for
only $\sim 10$ ms
\citep{shi06}.  Even in the case of NS/BH coalescence where the NS is
shredded, activity is not expected to last for more than a few seconds
\citep{dav05}.

In the model proposed here, the disruption of the companion star during 
coalescence could produce flares long after the SGRB. In the case of AIC,
by contrast, if the stellar companion could survive
the collapse of the NS to a BH and the accompanying impact of
neutrinos and gravitational radiation, then the
system may relax to a BHB system. Depending on the type of companion
star at the time of collapse, either a LMBHB or HMBHB will be
produced.  This scenario offers an alternate pathway for BHB and BH
microquasar formation.

Gravitational radiation is emitted in this model through two
mechanisms. One is the collapse of the NS to the BH. The collapse of
deformed, rapidly rotating NSs with millisecond periods would enhance
the gravitational wave output compared to collapse events that are
nearly spherically symmetric.  The coalescence of white dwarf cores
with masses of a few tenths of a $M_\odot$ onto NSs could also be
detectable, and we suggest that template waveforms for such events be
calculated for comparison with LIGO data.

\acknowledgments
We thank D.\ A.\ Kann, E.\ Nakar, P.\ S.\ Ray, and N.\ E.\ White for 
comments, and the referee for valuable suggestions.  The work of C.\
D.\ D.\ is supported by the Office of Naval Research.  The visit of
A.\ A.\ to the NRL High Energy Space Environment Branch to study this
problem was supported by NASA {\it GLAST} Science Investigation No.\
DPR-S-1563-Y, and a NASA Swift Guest Investigator Grant.

\end{document}